\documentclass[runningheads]{llncs}
\usepackage{graphicx}
\usepackage{hyperref}
\usepackage{xcolor}
\usepackage{amsmath}
\usepackage{tabularx}
\newcommand\blfootnote[1]{%
  \begingroup
  \renewcommand\thefootnote{}\footnote{#1}%
  \addtocounter{footnote}{-1}%
  \endgroup
}

\begin{document}

\author{Maisie Badami\inst{1}\and
        Marcos Baez\inst{2} \and
        Shayan Zamanirad\inst{1} \and
        Wei Kang\inst{3} }

\authorrunning{Badami et al.}

\institute{University of New South Wales (UNSW), Australia 
\email{\{m.badami,shayan.zamanirad\}@unsw.edu.au}\\
LIRIS – University of Claude Bernard Lyon 1, Villeurbanne, France \\
\email{marcos.baez@liris.cnrs.fr}\\
          Data61, CSIRO, Australia\\
\email{wei.kang@data61.csiro.au} }
\title{On how Cognitive Computing will plan your next Systematic Review} 

\maketitle

\begin {abstract}
Systematic literature reviews (SLRs) are at the heart of evidence-based research, setting the foundation for future research and practice. However, producing good quality timely contributions is a challenging and highly cognitive endeavor, which has lately motivated the exploration of automation and support in the SLR process. 
In this paper we address an often overlooked phase in this process, that of planning literature reviews, and explore under the lenses of cognitive process augmentation how to overcome its most salient challenges. In doing so, we report on the insights from 24 SLR authors on planning practices, its challenges as well as feedback on support strategies inspired by recent advances in cognitive computing. We frame our findings under the cognitive augmentation framework, and report on a prototype implementation and evaluation focusing on further informing the technical feasibility. 

\end{abstract}

\section{Introduction}
Systematic Literature Reviews (SLRs) are valuable research contributions that follow a well-known, comprehensive, and transparent research methodology. It is at the heart of evidence-based research, allowing researchers to systematically collect and integrate empirical evidence regarding research questions. Given their demonstrated value, SLRs are becoming an increasingly popular type of publication in different disciplines, from medicine to software engineering~\cite{kitchenham2004procedures}.
\blfootnote{This is a post-peer-review, pre-copyedit version of an article accepted to the International Workshop on AI-enabled Process Automation, at ICSOC 2020.}

Despite the valuable contributions of systematic reviews to science, producing good quality timely reviews is a challenging endeavor. Studies have shown that SLRs might fail to provide a good and complete coverage of existing evidence, missing up to 40\% of relevant papers~\cite{crequit2016wasted}, and even end up being outdated by the time of publication~\cite{sampson2008systematic,crequit2016wasted} -- this without considering those never published. 
\begin{figure}
\centering
\includegraphics[width=\textwidth]{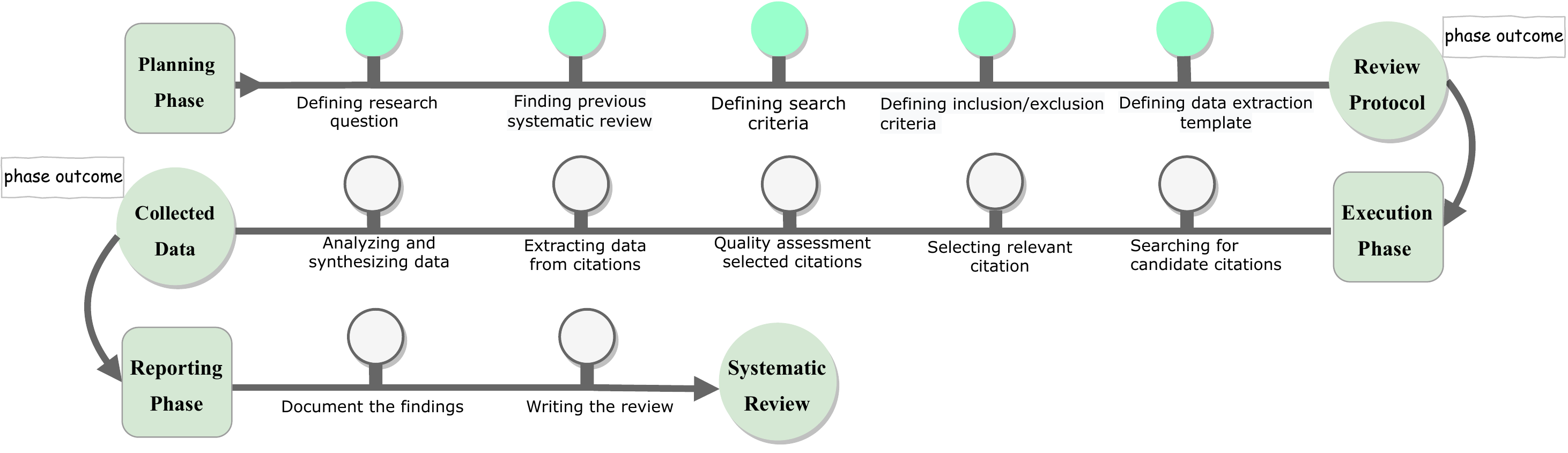}
\caption{The SLR process, defined by planning, execution and reporting activities} \label{fig:process}
\end{figure}

The reasons behind these challenges have been documented in several studies~\cite{hassler2016identification,garousi2017experience,palomino2018methodologies}, which attribute them to the demanding nature of the involved tasks, lack of expertise, limitations of support technology, and issues with primary studies.
Recent advances in cognitive computing and collaborative technology offer an opportunity to address these challenges, and support researchers in planning, running and reporting SLRs (see Figure \ref{fig:process} for an overview of the SLR process). We have seen new techniques and platforms enabling large-scale collaboration \cite{vaish2017crowd,sun2016crowdsourcing,krivosheev2018combining}, and automation opportunities \cite{howard2016swift,ouzzani2016rayyan,marshall2016robotreviewer}, offering promising results in different research activities relevant to the SLR process. Most of these efforts however are centered around the screening and identification of relevant scientific articles -- and rightly so as it is one of the most time-consuming phases -- but leaving other critical tasks largely unexplored.  

In this paper we address a much less explored phase of the SLR process, that of planning the reviews. Guidelines and recommendations (e.g.,~\cite{kitchenham2004procedures}) define the main activities in this phase as i) identifying the need for undertaking the review, ii) defining the research questions (RQs), iii) defining the search and eligibility criteria, and iv) the data extraction template. These tasks are fundamental to guiding the SLR process and setting the foundation to having meaningful and original contributions, good coverage of the literature and a process free of bias. Yet, as we will see, they are often poorly performed, if at all. 

In what follows we investigate how cognitive augmentation can support the planning phase of SLRs. We build on the insights and feedback from SLR authors to identify challenges and support strategies inspired by recent advancements in cognitive computing, framing the results under the framework for cognitive process augmentation. We also report on our early prototype and evaluation runs, showing the potential of augmentation in identifying the need for undertaking a review by leveraging word embeddings to find relevant SLRs from an input RQs.

\section{Challenges in Planning SLRs} 
The challenges in running an SLR can be found throughout the entire process. These have been observed in the literature \cite{hassler2016identification,garousi2017experience,palomino2018methodologies} as well as in our preliminary work, where we run an open-ended survey with more than 50 SLR authors tapping on their experience running SLRs. The results indicated that planning tasks are generally perceived as difficult to manage, requiring higher level of expertise and domain knowledge compared to more labor-intensive tasks.

\begin{figure}
\includegraphics[width=\textwidth]{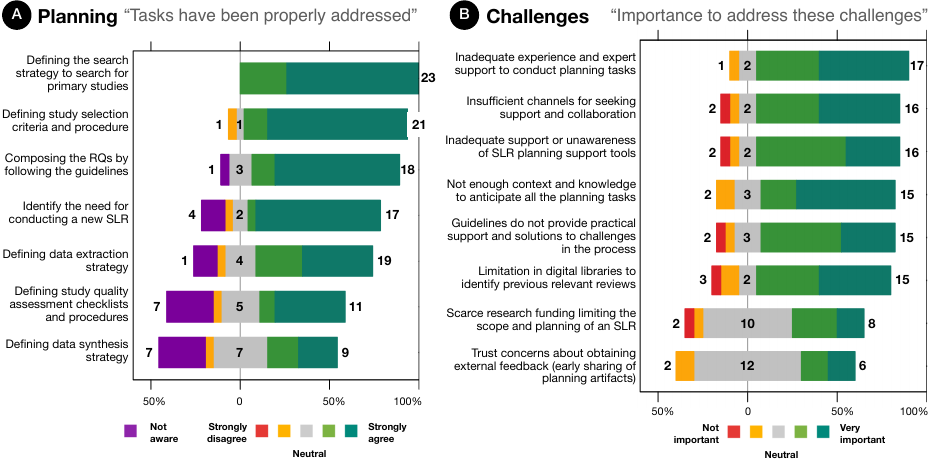}
\caption{Summary of feedback from SLR authors. A) Feedback on how planning tasks are addressed. B) Importance of addressing salient challenges in the planning.} \label{fig:Survey}
\end{figure}

Motivated by these insights, we run a second survey with 24 authors who published SLRs in top software engineering outlets in the last two years. This survey focused on their experience in planning SLRs, and inquired about i) whether planning tasks were properly addressed in their last SLR project, ii) the importance of addressing some salient challenges, and asked for iii) feedback on some support strategies to address the emerging challenges.  In Figure \ref{fig:Survey} we summarise the feedback on the first two points.

While most authors reported positively to having addressed the tasks properly -- not surprisingly given the quality of the outlets -- there was still a significant number of researchers reporting neutral to negative, and in some cases even not being aware of certain tasks. More illuminating is to observe the challenges that we identified and the importance authors put in addressing them (Figure \ref{fig:Survey}), which we summarise and group below and address in the next section:

\begin{itemize}
    \item[\textbf{C1}] Inadequate experience and support by current tools and guidelines.
    
    \item[\textbf{C2}] Insufficient context and knowledge to anticipate  tasks in the planning phase. 
    
    \item[\textbf{C3}] Limitations of digital libraries to identify relevant SLRs. 
    
    \item[\textbf{C4}]  Inadequate expert support and trust concerns in obtaining external feedback.
    
\end{itemize}
 
\section{Cognitive Support in Planning SLRs}

In this section, we present the conceptual design of a platform to support the planning of SLRs. 
In the following we describe the solutions and strategies addressing the main challenges, and the feedback obtained from SLR authors.

\noindent \textbf{Strategies to address planning challenges}.
 The strategies were derived in brainstorming sessions among the co-authors, taking as input the insights from the first round of interviews with SLR authors on workarounds and strategies employed in the process, our own experience and prior work. The resulting strategies leverage techniques from machine learning and data-mining to address the main challenges.
Below we describe the strategies for each of the challenges (C\#).

\begin{itemize}
\item \textbf{Chatbot assistant} that allows authors to ask questions about the SLR process and best practices (C1). The chatbot provides a natural language interface to all information encoded in guidelines and recommendations.  
\item \textbf{Step-by-step guides} for each of the tasks (C1). The interface provides a conversational interface that assists authors in the preparation of each “artefact”, (e.g., RQs), by providing step-by-step prompts based on guidelines.
\item \textbf{Incremental and iterative process} that can be adapted as more information becomes available to the researcher (C2). For example, RQs and inclusion/exclusion criteria can be refined as we identify similar literature reviews and learn more about the topic. 
\item \textbf{Incorporating context and knowledge} (C2). The system leverages information available in seed papers and similar literature reviews by extracting SLR-specific metadata relevant to the protocol (e.g., RQs, search strategy) and making them available to the authors as a reference point at each step.  
\item \textbf{Search focused on similar SLRs} (C3). Instead of defining complex queries and terms to identify similar reviews, the search focuses only on SLRs. The search results provide SLR-specific information, including RQs, inclusion/exclusion criteria, search strategy, etc.
\item \textbf{From RQs to similar SLRs} and papers (C3). The system allows users to go from their (partial) RQs directly to similar SLRs by enriching information with extra data from seed papers.
\item \textbf{Expert networking} (C4). We presented strategies including artefact-specific evaluation criteria, improved discovery of experts, and leveraging groups of trust. However, we limit the discussion on human-human collaboration.
\end{itemize}

\begin{figure*}[t]
\includegraphics[width=\textwidth]{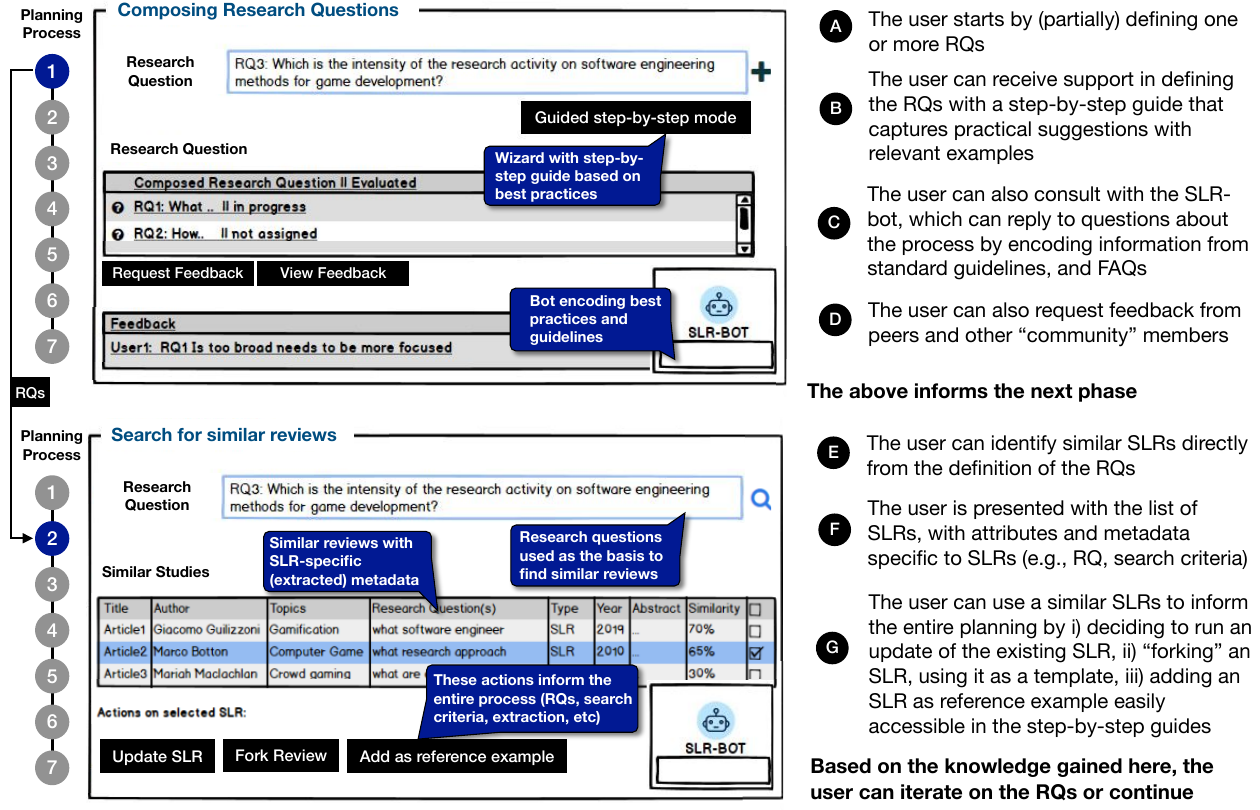}
\caption{Interaction illustrating the first two steps in the planning: defining the RQs and identifying the need for the SLR. The same strategies apply to the rest of the steps.} 
\label{fig:Concept}
\end{figure*}

We illustrate how the above strategies can be combined in a the concept tool in Figure  \ref{fig:Concept}, depicting the iteration over the first two steps of the planning.

\noindent \textbf{Feedback from SLR authors}. We requested feedback on the concept tool and strategies in our second wave of survey with 24 SLR authors. Authors were shown a mockup and descriptions,\footnote{All materials related to the study can be found at~\url{https://github.com/maisieb01/Cognitive_SLR.git}}
, and asked to i) rate the strategies on whether they addressed the specific challenges, ii) provide feedback on potential barriers for adopting them, as well as iii) other alternatives not foreseen.

The strategies to address the inadequate experience and support (C1) received positive feedback by almost all participants (15/16). For example, a participant stated \textit{``I had to search for hours to understand how to get tasks done, so I think a Bot that helps with the common question would be really helpful"}. Another one  stated \textit{``Support from peers is the most important help that I've been missing when conducting my study. [..] I didn't know much about the process and where to start, the step by step guideline would have helped me a lot"}. 

Participants reported potential barriers being, i) mismatch between the assistance provided by the tool and practices of the target domain of the SLR, and ii) the quality of the underlying algorithms, including the actual bot recommendations.

Regarding the lack of context and knowledge (C2), all participants provided positive feedback (17/17). The only concern was about extracting information from multiple SLRs following different approaches and the tool mixing them up.

The strategies to address the limitation of digital libraries (C3) also received positive feedback (15/17). The argument against it came from an author not sure about the effectiveness \textit{``I'm not sure if searching by [RQs] will help identify an area which needs more research or just other SLRs .. [it] just seems like a `type' filter as e.g. in Scopus or adding `review' in Title from Google Scholar"}.

Among the suggestions we can mention: i) adapt the builtin guidelines to the target domain of the SLR, and consider other frameworks such as PICO, ii) extend the SLR extraction capabilities to recommend highly reputable venues, infer the guidelines to follow and create a to-do list authors could follow, iii) expand the search to suggest research that is very relevant, and iv) include explanations for as to why papers are suggested.

\section{Conceptual Architecture}
In delivering on the vision of cognitive support in the planning of SLRs, we rely on the framework for cognitive process augmentation by Barukh et al.~\cite{barukh2020}. 

As we will see, the strategies we devised inform this architecture at different layers.

\indent\textbf{Foundation:} 
Existing technology provides support for coordinating \textit{data}, \textit{tasks} and \textit{collaboration} that we can leverage to build our vision of a cognitively-augmented planning process. 
Starting from the process itself, we have seen the planning of SLRs to be an incremental and iterative process leading to a review protocol. The process management in this context could rely on lightweight artifact-centric systems (e.g., Gelee~\cite{bssez2009gelee}) where the researcher drives the process while the system advises on the steps to take based on community-specific guidelines. 
Along the process, some tasks are already supported by current online services, such the search and access to scientific articles. Digital libraries and search engines provide access to article data and metadata, but under the limitations pointed out in the previous section, requiring researchers to engage in significant manual effort. Thus, although the data and knowledge required to elaborate the research protocol and inform the process is available, identifying, curating and adopting such knowledge is a challenging endeavour.

\indent\textbf{Enablement:} 
The next layer leverages existing data sources and services to apply domain-specific data extraction and enrichment that will enable cognitive augmentation. Components such as \textit{article recommendation}, enabling search for similar SLRs and papers, \textit{article augmentation}, enriching SLRs with domain-specific metadata, \textit{activity recommendation}, recommending steps based on process definitions and progress, and \textit{knowledge graph}, aggregating knowledge about the process in queryable format, are among the enabling components. 
In this context, SLRs, primary research articles, and guidelines on how to run the SLR process, are the main sources of information. Lower-level algorithms such as named-entity recognition, word-embeddings and similarity serve as building blocks for these higher level components. 

\indent\textbf{Delivery:} 
The researcher finally experiences the cognitive augmentation in the planning through Conversational AI as well as intelligent GUIs. Conversational AI helps in delivering assistance in the process, providing a natural language interface to query the vast knowledge encoded in guidelines, and receive practical assistance in each step of the process. In the form of more general conversational interfaces, guided prompts would provide step-by-step guidance to assist researchers in knowledge-intensive tasks (e.g., defining RQs). We also recognise the need for serving more traditional delivery systems such as GUIs, to dote complex tasks with intelligent features (e.g., domain-specific search).

\section{Prototype Implementation and Evaluation}
In this section we describe our ongoing exploration into the technical feasibility of our approach.

We started with \textit{article recommendation} as it emerged as a promising component based on the feedback from SLR authors, but raised concerns in terms of feasibility. We note however that providing support for identifying related SLRs has all  the potential benefits but fewer of the concerns (e.g., recall and accuracy) with respect to assisting the identification and screening of relevant literature~\cite{kontonatsios2017semi}. In the planning, bringing up the most relevant SLRs and papers can provide the additional context in a task that some people are not even aware of.

\begin{figure*} 
\centering
\includegraphics[width=\textwidth]{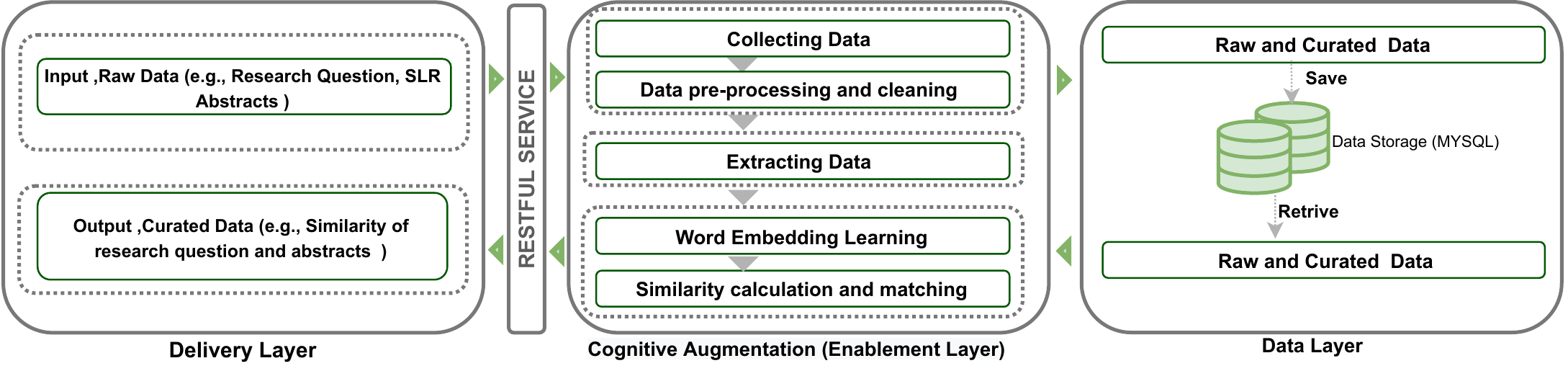}
\caption{Architecture Pipeline} \label{fig:Pipeline}
\end{figure*}

\indent\textbf{Prototype}. The prototype is our initial exercise into understanding the technical requirements of the system, as well as a working tool to serve evaluation purposes. On the surface, the current version of the prototype provides a set of REST APIs (and an accompanying user interface) that given a set of RQs in input, it returns the most relevant SLRs, along with the relevance score computed based on the available models. 
Fig.~\ref{fig:Pipeline} presents the pipeline of the implementation.

The source of information is currently a curated database of SLRs, where domain-specific information has been manually extracted so as to evaluate the recommendation component in isolation. The end goal is to have a data layer that can interface with existing services to access structured and unstructured data, which can then be processed to automatically extract relevant domain-specific information. 
While we currently use a MySQL database to store and retrieve raw and curated data, the concept of \textit{Data Lake} \cite{beheshti2017coredb} emerges as a promising direction to store and query structured and unstructured SLR data. 

In entering the augmentation layer, the data is pre-processed in two steps: (a) normalizing text corpus (e.g., removing special characters and stop words, all to lowercase); and (b) lemmatizing and converting each word to its base form. We leverage Stanford CoreNLP toolkit\footnote{https://nlp.stanford.edu/software/tagger.html} to perform this process.

Following the data cleaning, the next step is to extract meaningful information from the data.
The inspiration behind our proposed approach is leveraging a similar approach to the word embedding's model~\cite{mikolov2013distributed} that represents words in a Vector Space Model (VSM). We have extended the idea of considering a "word" as a vector to represent the SLR-related corpus (e.g., RQs and SLR abstracts) in a vector space. To create the VSMs, we employ an N-gram selector component to extract all the keywords (nouns and verbs) from sentences of the given context. We leverage Stanford Part-Of-Speech (POS) Tagger~\cite{manning2014stanford} to achieve this. 
Then, we create a list of n-grams out of these keywords and transform them into vector representations for each corpus. After encoding the given corpora into vectors, these vectors are used to calculate the similarity between desired corpora.
Information is then augmented based on a pool of word embedding models. Our work leverages state-of-the-art algorithms widely used in NLP communities. Several such algorithms (e.g. GloVe~\cite{pennington2014glove}, Word2Vec~\cite{mikolov2013efficient}, Numberbatch~\cite{speer2016conceptnet}, WikiNewsFast~\cite{bojanowski2017enriching}, and GoogleNews~\cite{church2017word2vec}) come with efficient implementations that are readily available as libraries to use.
The REST APIs then expose the functionality of the article recommendation component for programmatic access. A front-end application takes these services and wraps them up in a user interface.

\indent\textbf{Planned Evaluation}.
The goal of the initial evaluation is to inform specific design decisions regarding the algorithmic support for recommending papers. Among the main design decisions we consider: i) \textit{What models will better serve the specific task?}
The idea is to identify among the embedding models and architectures the most promising candidates to build on, and understand whether investing in domain-specific embedding models is required. Then, ii) \textit{What information should we leverage when assessing the relevance calculations?} (e.g., title, title-abstract, RQs, full-text). The aim is to understand what (combinations of) information to focus when assessing the relevance of SLRs from an input RQs, and therefore to consider in the extraction process. 

The prototype supports these two dimensions, models and selective information, so as to serve the evaluations. The dataset of SLRs is being manually constructed to incorporate for each SLR a set of related SLRs (as reported in the reviews) and not relevant SLRs as judged by human experts. 
Armed with the human-annotated dataset, we evaluate the quality of word embedding models by assessing how well the similarity scores of the word vectors correlate with human judgment~\cite{shi2017jointly}. The similarity is calculated as the distance between the vectors representing RQs and SLRs, using cosine similarity as measure, which has been found suitable for SLRs in prior work \cite{lopes2007visual}.

We rely on Spearman’s rank $(r_s)$ correlation between the word embedding models similarity score and researchers annotations to evaluate how well the similarity of the given pairs (e.g., RQs and abstracts) agrees with human judgments~\cite{huang2012improving}.
We performed a preliminary test run to tune the experimental setup, and the early results are encouraging. Results under limited settings (e.g., dataset of 160 SLRs, only abstract-RQ comparisons) already show good level of agreement ($r_s$ = 0.67), for the best performing model, although at this point this is anecdotal since more comprehensive tests are required. 

\section{Conclusion and Future Work}
We have seen that planning is a challenging endeavor, requiring resources, expertise or context that is often missing when undertaking a new topic, when performing it for the first time, or when resources are lacking -- all typical scenarios in research. This paper shows that cognitive processes provide the ingredients to address these issues and support researchers in this often overlooked but impactful phase. 
As for ongoing and future work, we are in the process of refining the technical details of the first experiment, and planning an evaluation with end-users so as to assess the actual benefits of the approach when compared to standard tools. We also continue with our human-centered design approach to the development of the overall tool and algorithms, which will inform all components of the platform.

\section{Acknowledgments} We acknowledge CSIRO Data61 for funding scholarship on this research.

\bibliography{references.bib}
\bibliographystyle{plain}

\end{document}